\RequirePackage{amsmath}
\documentclass[runningheads]{llncs}

\usepackage[T1]{fontenc}
\usepackage{graphicx}
\usepackage{amssymb}
\usepackage{xcolor}
\usepackage{xurl}
\usepackage{hyperref}
\hypersetup{
    colorlinks=true,    
    linkcolor=black,    
    citecolor=blue,     
    hidelinks           
}
\usepackage{csquotes}
\usepackage[capitalize]{cleveref}
\usepackage{adjustbox}
\usepackage{caption}
\usepackage{booktabs}
\usepackage{multirow}
\usepackage{tikz}
\usetikzlibrary{shapes, arrows, shapes.arrows,arrows.meta}
\definecolor{itdcolor}{RGB}{185,164,198}

\newcommand{\nR}{\mathbb{R}}

\DeclareMathOperator{\avg}{avg}

\usepackage[style=nejm,
citestyle=numeric-comp,
sorting=none]{biblatex}
\newif\ifBlindReview
\BlindReviewfalse

\newif\ifNotBlindReview
\ifBlindReview\NotBlindReviewfalse\else\NotBlindReviewtrue\fi

\newif\ifArXiv
\ArXivtrue

\newif\ifNotArXiv
\ifArXiv\NotArXivfalse\else\NotArXivtrue\fi

\newif\ifShowComments
\ShowCommentsfalse

\begin{document}

\title{Topology-driven identification of repetitions in multi-variate time series}

\ifBlindReview
	\author{Anonymous Author(s)}
	\authorrunning{Anonymous}
	\institute{Anonymous Institution\\
		Anonymous Location\\
		\email{anonymous@example.com}}
\else
	\author{Simon Schindler\orcidID{0000-0003-3050-4456} \and
		Elias Steffen Reich\orcidID{0009-0000-4561-9067} \and \\
		Saverio Messineo\orcidID{0000-0003-1592-4428} \and
		Simon Hoher\orcidID{0000-0001-8415-7520} \and \\
		Stefan Huber\orcidID{0000-0002-8871-5814}}
	\authorrunning{S. Schindler et al.}
	%
	\institute{Josef Ressel Centre for Intelligent and Secure Industrial Automation\\
		Salzburg University of Applied Sciences, Austria \\
		\email{\{simon.schindler,eliassteffen.reich,\\saverio.messineo,simon.hoher,stefan.huber\}@fh-salzburg.ac.at}}
\fi
\maketitle              
\begin{abstract}
	Many multi-variate time series obtained in the natural sciences and engineering
	possess a repetitive behavior, as for instance state-space trajectories of industrial machines in discrete automation. Recovering the times of recurrence from such a multi-variate time series is of a fundamental importance for many monitoring and control tasks.
	For a periodic time series this is equivalent to determining its period length.
	In this work we present a persistent homology framework to estimate recurrence times in multi-variate time series with different generalizations of cyclic behavior (periodic, repetitive, and recurring). To this end, we provide three specialized methods within our framework that are provably stable and validate them using real-world data, including a new benchmark dataset from an injection molding machine.
\end{abstract}

\section{Introduction}

\subsection{Motivation}
Determining when a system returns to a previous state is of high importance in
many fields of science and engineering. In industrial automation, for example,
the state-space trajectory of a machine can provide valuable insights into its
operation. Even though this is a fundamental problem that should be addressed
for improving tasks such as anomaly detection, the industry currently lacks a robust and
general method to identify
repetitive behavior in multi-variate time series~\cite{Mueller2025}. By
identifying the time points at which the machine returns to a specific
state, one can evaluate the system on the basis of production cycles and detect
anomalies or wear in the machine components.
The identification of repetitive behavior in multi-variate time series is also
relevant in other fields, such as biology, where the cyclic behavior of
biological systems can provide insights into their underlying mechanisms such
as circadian rhythms.

In this work, we address the problem of accurately estimating the times
at which a multi-variate time series returns to a specific state, even in the
presence of noise and non-uniform sampling conditions.
To this end, we propose a framework based on persistent homology
and three applicable methods each tailored
to a specific type of repetitive behavior: periodic, repetitive, and recurring.

\subsection{Problem setting}

\paragraph{\textbf{Basic definitions.}}\label{sec:definitions}

Considering a multi-variate time series $x$ over $n$ real variables over a
finite timespan $I = [0, T] \subset \nR$, we formalize $x$ as a parametrized curve $x
	\colon I \to \nR^m$, which must not necessarily be continuous.\footnote{Although the time series need not be continuous, it must have a finite number of discontinuities.}
For the following we endow $\nR^m$ with the Euclidean norm $\|\cdot\|_2$, although one could also consider different metric spaces.
As we are interested in the \enquote{cyclic behavior} of
$x$, we will consider \enquote{cyclic behavior} in three different levels of
strictness in the following.
In all cases we decompose the timespan $I$ into $k$ intervals
$I_i = [T_{i-1}, T_i]$ with $0 = T_0 < T_1 < \dots < T_k = T$ such that each
$I_i$ corresponds to one such \enquote{cyclic iteration}. We define $\tau_i = |
	I_i| = T_i - T_{i-1}$ as the \enquote{cycle length} of the $i$-th iteration.

What follows is a definition of the three notions of \enquote{cyclic behavior}
we consider in this paper starting with the most strict notion:

\begin{definition}[periodic]
	We call $x$ to be \emph{periodic} with period length $\tau \in (0, T)$ if
	$x(t+\tau) = x(t)$ for all $t \in [0, T-\tau]$.
\end{definition}

Consequently, $T_i = i \tau$ for $i < k$ for a periodic time series $x$.
Figuratively speaking, we
refer to $x$ as repetitive if $x$ is periodic with possibly varying periods and up to
some reparameterization per period.

\begin{definition}[repetitive]
	We call $x$ to be \emph{repetitive} if there are $T_0, \dots, T_k$ as
	mentioned and continuous maps $\gamma_i: I_i \to I_1$, which are non decreasing with $\gamma_i(T_{i-1}) = T_0$, and
	for $i < k$ also surjective, such that
	$x(t) = x(\gamma_i(t))$ for all $t \in I_i$.
\end{definition}

If also $\gamma_k$ is surjective then also the last repetition is
\enquote{complete}, which we do not require in general.
A $\tau$-periodic $x$ is also $m\tau$-periodic for all $m \in \mathbb{N}$. To resolve
this ambiguity, we refer to the smallest possible period when we speak of the
period of a time series, and analogous for repetitive time series.

\begin{definition}[recurring]
	If $\{T_0, \dots, T_k\} = \ker(x-x(0)) \cup \{T\}$ and in \newline particular $\ker(x-x(0)) = \{t \in I \colon x(t) = x(0)\}$ is finite, we call $x$ \emph{recurring}.
\end{definition}

We are of course interested in the non-trivial cases of recurring
multi-variate time series, i.e., when $k > 0$. Note, if $x(T) \neq x(0)$ we
still want $T_k = T$, which is formally reached by considering
$\ker(x-x(0)) \cup \{T\}$ in the definition.

Time series from real-world measurements typically possess some sort of noise
or other forms of variations. Hence, we also define $x$ to be
approximately periodic, repetitive or recurring if it adheres to the following:

\begin{definition}[approximately I]
	\label{def:approx1}
	The time series $x$ is \emph{$\varepsilon$-approximately periodic
		(repetitive, resp.)} if there is a multi-variate time series $\hat{x}$ with
	$\|\hat{x} - x\|_\infty \leq \varepsilon$ that is periodic (repetitive, resp.).
\end{definition}

Applying this definition also to recurring $x$ bears the problem of noise
leading to many tiny cycle iterations, i.e., where the farthest point of a cycle
iteration is only arbitrarily closer to $x(0)$ than $\varepsilon$.
We, therefore, also add a condition on the required significance of
cycle iterations in terms of how far the farthest point needs to be away from
the closest point to \enquote{create} a cycle.

\begin{definition}[approximately II]
	\label{def:approx2}
	The time series $x$ is \emph{$\varepsilon$-$\delta$-approximately
		recurring} if there are $0 = T_0 < \cdots < T_k = T$ such that for all $i
		\in \{1, \dots, k-1\}$
	\begin{itemize}

		\item $x(T_i)$ is a locally closest point to $x(0)$ with $\|x(T_i) - x(0)\|
			      \leq \varepsilon$, and

		\item for all $j \in \{i, i+1\}$, for $x(t)$ being the farthest point of
		      $x$ from $x(0)$ over $I_j$, we have $\|x(t) - x(0)\| > \delta + \|x(T_i)
			      - x(0)\|$.

	\end{itemize}
\end{definition}

In practice, we think of $\varepsilon$ as being just as small to account for
noise, and $\delta$ as being just as large to account for the
\enquote{diameter} of a cycle in terms of distance from $x(0)$. Also note,
if $x$ is $\varepsilon$-$\delta$-approximately recurring then there is a
recurring $\hat{x}$ with $\|x - \hat{x}\|_\infty < \varepsilon$. In this sense, \cref{def:approx2} is stricter than \cref{def:approx1}.

\paragraph{\textbf{Problem statement.}}

In this paper we are concerned with finding all $\tau_i$, or equivalently, all
$T_i$ for approximate periodic, repetitive and recurring multi-variate time
series.
We assume the multi-variate time series is given through $n$ samples,
i.e., a finite series $x(t_1), \dots, x(t_n)$, at time stamps $t_1 < \dots <
	t_n$. The time stamps may not necessarily be equidistantly spaced.

Note, often in multi-variate time series analysis, a time series $x$ is
decomposed as a sum $x_t + x_r$ of a trend signal $x_t$ and a residual signal
$x_r$. In this case our investigation addresses the residual signal $x_r$. In
the remainder of this paper we may assume $x$ to have no trend.

\subsection{Prior and related Work}

\paragraph{\textbf{Limitations of classical periodicity detection methods.}} The methodology proposed in this paper can handle unevenly
spaced time samples and varying period lengths, unlike classical periodicity
detection methods. These classical approaches fall into three categories:
frequency-domain, time domain, or hybrid methods. Frequency-domain techniques
use Fourier transform, typically implemented via FFT, which requires evenly
spaced samples. Similarly, the time domain auto correlation function fails with
uneven sampling since correlation becomes ill-defined~\cite{White2001}. The
wavelet transform, a hybrid approach implemented through filter banks in its
discrete form, also requires evenly spaced samples~\cite{White2001}.
Furthermore, these methods can only detect the overall periodicity of a time
series, not individual cycle lengths.
Thus, classical methods cannot address even the simplest case considered in this paper: periodic time series with uneven sampling.

\paragraph{\textbf{Topology of time series.}} This paper is not the first to exploit the topological structure of time
series to extract information.
In~\cite{Perea2015,Perea2015a,Perea2016,Perea2019}, Perea et al.\ examine the characteristics of
delay embeddings applied to (quasi-)periodic functions, discovering
they form loops without self intersections in the embedding space when the embedding dimension is
sufficiently high. However, unlike our approach, they focus solely on
periodic and quasi-periodic functions and do not address the recovery of
period lengths.

In~\cite{Perea2018}, they propose a technique to quantify the
(quasi-)periodicity of video data, which is a type of multi-variate time
series. For this purpose, they estimate the period length to determine the
minimal sufficient delay embedding dimension. This involves constructing
a 1D surrogate series using the first coordinate of a diffusion map
of the video frames and employing an autocorrelation-based method to
estimate the period.

Yang et al.~\cite{Yang2016} follow a related strategy for period estimation
in video data. Their surrogate series is constructed by measuring the
mutual information between the first frame and subsequent frames. They
then apply a heuristic peak detection algorithm to estimate the period
length. Their method for peak detection does not come with any stability
guarantees.

Note, the above methods are not directly applicable to our problem, since
they are focused on periodic and quasi-periodic functions, with constant period
length, and not on repetitive/recurring functions.

In~\cite{Bonis2024}, Bonis et al. focus on analyzing
multiple periods of a series that have been transformed through a reparameterization
\(y_{\text{observed}}(t) = f(\gamma(t))\). Their objective is to determine \(\gamma^{-1}\)
and identify the number of periods present in the series. They achieve this by analyzing points
in the persistence diagram resulting from the sub-levelset filtration of the series.
Their study diverges from ours in two significant ways: firstly, their goal is to determine
the number of periods rather than the lengths of these periods. Secondly, their analysis
does not extend to multi-variate series.

In~\cite{Bauer2024}, Bauer et al.\ propose a method embedding a multi-variate time series
along with its first normalized numerical derivative (representing direction) into a
higher-dimensional space. They then analyze subsequences of the embedded time series by
investigating the H1 persistence of the Rips filtration of the point cloud generated
from these subsequences to determine the number of cycles present. They do not explicitly
extract the period length of the time series. Note, the use of the first
numerical derivative of the time series makes their method more susceptible to noise
compared to ours. Moreover, in our method direction is also implicitly encoded through the use of
time delay embeddings.

Instead of considering the topology of the state space trajectory of a time series directly
(or of its delay embedding), in~\cite{Ichinomiya2023} Ichinomiya et al.\ conclude a super- and
sublevel set filtration of the recurrence function of a multi-variate time series, to construct
topological features useful to study dynamical systems. However, they do not extract
the period length of the time series.

\paragraph{\textbf{Summary of the state of the art.}}

Besides the work of Perea et al.\ in~\cite{Perea2018} and by Yang et al.
in~\cite{Yang2016}, there is no work directly addressing the problem of
identifying the period length of a multi-variate time series. The methods
proposed by Perea et al.\ and Yang et al.\ are not applicable to our
problem, since they are focused on (quasi-)periodic videos, i.e.\ state
space trajectories, with periodic, and not repetitive/recurring
behavior.

The literature also lacks high-quality real-world benchmark datasets with annotated recurrence times validating such methods.

\subsection{Contribution}

The main contributions of this paper are:
\begin{itemize}
	\item The introduction of a computationally persistent homology-based framework for identifying recurrence times in multivariate time series, providing a versatile and general approach going beyond specific domains or signal types, thus filling a gap in the literature.
	\item The presentation of three methods to use the proposed framework suited to handle periodic, repetitive, and recurring behavior and come with stability guarantees under perturbations.
	\item The proposal of a real-world, high-fidelity benchmark dataset containing multi-variate time series from an injection molding machine, specifically designed to reveal periodic, repetitive, and recurring behavioral patterns and to validate the performance of the proposed methods within our framework.
\end{itemize}

\section{Background}
\subsection{Persistent homology}
An essential aspect of the methods introduced in this work is the detection and
quantification of significance of local minima of a continuous
scalar-valued function defined on a bounded interval. For this purpose, we
apply a sublevel set filtration of the function of interest to aggregate
information about its topology into a so-called persistence diagram. The
following provides a comprehensive introduction to the necessary background on
zero-dimensional persistent homology since higher-dimensional persistent homology
is not required for the methods presented in this paper.
Please refer to~\cite{Huber2020} for a
more in-depth treatment.

Given a continuous function $f: I \to [0,\infty)$ defined on a bounded interval $I =
	[0, T]$ possessing a finite number of critical points, we construct the
sublevel set filtration of $f$ as a collection of increasing spaces
$S_\lambda$, where $S_\lambda = f^{-1}([0, \lambda])$ are the sublevel sets of $\lambda$ for $\lambda \in [0,\infty)$.
We obtain the persistence diagram $D(f)$ by tracking the appearance and
disappearance of connected components within the sublevel set filtration of
$f$. We refer to the connected components of $S_\lambda$ as the compact subsets of $S_\lambda$ of maximum length.

Consider the set $M$ of local minima of $f$.
In the sublevel set filtration a local minimum $t \in M$ is born / first
appears in $S_b$ at time $b = f(t)$. It then dies / disappears at time $d$,
when the connected component containing $t$ merges with another connected
component in $S_d$.

By tracking the birth and death times of local minima, we can construct the
persistence diagram $D(f)$, which consists of a (multi)-set of points in the
plane. Each point $(b,d)$ in $D(f)$ corresponds to one or multiple local minima appearing at time $b$ and disappear at time $d$. We record this
correspondence $M \to D(f)$ while constructing the sublevel set filtration of
$f$.
The persistence of a point $(b,d)$ is defined as $d-b$, which quantifies the
significance of the corresponding local minima.

We also include the set of points with zero persistence in $D(f)$, with each
point $(b,b)$ having infinite multiplicity.
This inclusion is important for the following property to hold: The persistence
diagram is stable under perturbations $f'$ of $f$ with respect to the
bottleneck distance $d_B(F, F')$, which is a metric between two persistence
diagrams $F = D(f)$ and $F' = D(f')$ in the space of persistence diagrams.
Considering all possible bijections $\phi: F \to F'$, the bottleneck distance
is defined as
\begin{align*}
	d_B(F, F') = \inf_{\phi} \sup_{p \in F} \|p - \phi(p)\|_\infty
\end{align*}
The bottleneck distance between the persistence diagrams of $f$ and it's perturbed
version $f'$ is bounded by the distance between $f$ and $f'$ in the
sup-norm~\cite{Cohen2005}:
\begin{align}\label{equ:stability_PD}
	d_B(D(f), D(f')) \leq \|f - f'\|_\infty
\end{align}
This property ensures the robustness of the persistence diagram under
perturbations of the function $f$, such as noise, and guarantees the overall
stability of our methods.

The runtime complexity of computing the persistence diagram of a discrete
scalar-valued function $f$ is $O(n \alpha(n)$, where $n$ is the number of
samples of $f$ and $\alpha$ is the inverse of the Ackermann function, which is
for all practical purposes bounded from above by a constant~\cite{Edelsbrunner2010}.
This makes persistent homology a computationally
efficient tool for analyzing time series data.

\section{Topology-driven identification of repetitions}\label{sec:methods}
In this section we introduce a general framework to estimate the
times of recurrence of multi-variate time series. The essence of our approach
is to construct a scalar-valued function $v$ from the multi-variate time series
and apply the sublevel set filtration to this function to extract topological
features $D(v)$. We then analyze these features to estimate the times of
recurrence. The strength of this approach lies in it's real-world
applicability, generality and flexibility. It only requires a stable surrogate
function $v: I \to [0,\infty)$ capturing the relative position within the
current cycle on the state space trajectory over the time interval $I$ to
construct the persistence diagram $D(v)$. This allows us to apply the same
method to a wide range of multi-variate time series exhibiting cyclic behavior.

To demonstrate the effectiveness and versatility of our approach, we present
three methods tailored to multi-variate time series exhibiting more constrained
cyclic behavior, i.e., recurring, repetitive, and periodic time series.

\begin{figure}
	\begin{minipage}{\textwidth}
		\centering
		\begin{tikzpicture}[scale=0.8, transform shape]
			\node[shape=rectangle,draw=black,line width=0.25mm,minimum size=0.1cm,fill=itdcolor!50] (x) at (0,0) {$x: I \rightarrow \mathbb{R}^m$};
			\node[shape=rectangle,draw=black,line width=0.25mm,minimum size=0.1cm,fill=itdcolor!50,xshift=1.5cm] (v) at (3,0) {$v_x: I \rightarrow [0,\infty )$};
				\node[shape=rectangle,draw=black,line width=0.25mm,minimum size=0.1cm,fill=itdcolor!50,xshift=1.5cm] (u) at (3,2) {$U_x: I \rightarrow [0,\infty )^{md}$};
			\node[shape=rectangle,draw=black,line width=0.25mm,minimum size=0.1cm,fill=itdcolor!50,xshift=1.5cm] (w) at (3,-2) {$w_x: I^2 \rightarrow [0,\infty )$};
			\node[shape=rectangle,draw=black,line width=0.25mm,minimum size=0.1cm,fill=itdcolor!50,right of = v,xshift=4.3cm] (p) at (6,0) {$D_{0}: \mathcal{L}^\infty(I,\mathbb{R}) \rightarrow \{(b_i,d_i)\} \subset \mathbb{R}^2$};
			\draw [->,line width=0.25mm](x) -- node[above]  {$L_2$ Distance} node[below]  {Method 1}  (v);
			\draw [->,line width=0.25mm](x) |- node[above,xshift=1.5cm]  {Delay Embedding} node[below,xshift=1.5cm]  {Method 2} (u);
			\draw [->,line width=0.25mm](x) |- node[above,xshift=1.5cm]  {$L_2$ Distance} node[below,xshift=1.5cm]  {Method 3} (w);
			\draw [->,line width=0.25mm](u) -- node[right]   {$L_2$ Distance}(v);
			\draw [->,line width=0.25mm](w) -- node[right]   {Average}(v);
			\draw [->,line width=0.25mm](v) -- node[above]  {Sublevel Set Filt.}(p);
		\end{tikzpicture}
		\caption{Overview of the function spaces in our approach of first constructing a scalar-valued function $v$ from a multi-variate time series $x$ with the three different methods presented and subsequently applying the sublevel set filtration to estimate the times of recurrence.\vspace{1cm}}
		\label{fig:method_overview}
	\end{minipage}
	\begin{minipage}{\textwidth}
		\centering
		\includegraphics[width=0.8\textwidth]{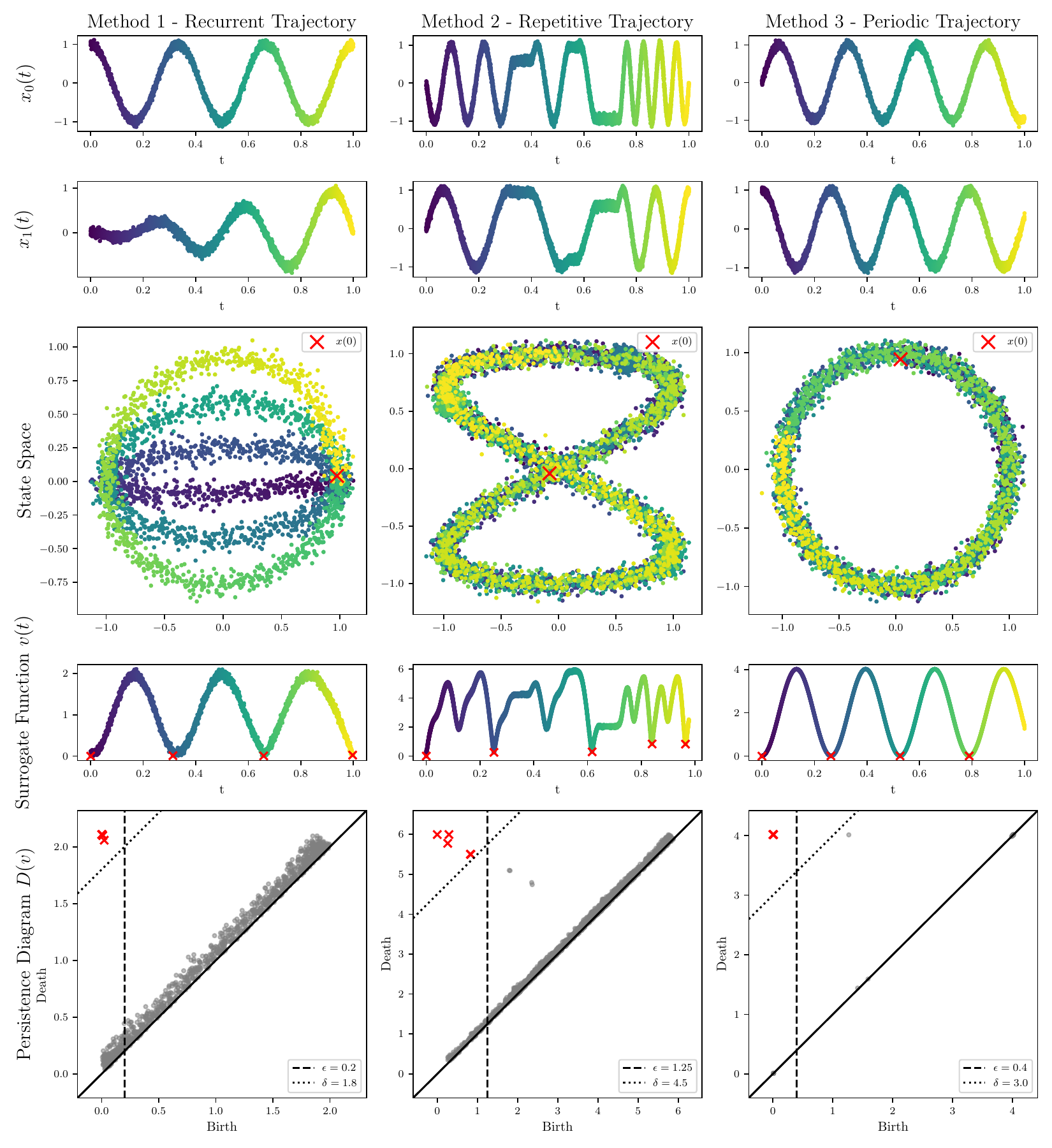}
		\caption{From left to right: recurrent, repetitive, and periodic time series, each with the corresponding scalar-valued function $v_x, v_x, v_{x}$ constructed by the three methods and the resulting persistence diagram $D(v_x), D(v_x), D(v_{x})$.}
		\label{fig:method_viz}
	\end{minipage}
\end{figure}

As the class of time series gets more constrained, we incorporate more
structure into our methods to exploit the specific characteristics. Method 1 is
of general nature and can be applied to any multi-variate time series, where
the state space trajectory is known to recur at least once to a point close to
the starting point. Method 2 is tailored to repetitive time series with
possible self-intersections in a single repetition and Method 3 is designed to
handle periodic time series. \ifArXiv We provide proofs of the stability of
	these methods under perturbations of the input time series in the
	\cref{sec:stability_proofs}.\fi
\cref{fig:method_overview} provides an overview of the three methods and the
general framework.

\subsection{Determination of Recurrences (Method 1)}
\paragraph{\textbf{Method.}} As introduced in \cref{sec:definitions}, we consider
a time series $x$ over a finite interval $I = [0, T]$ to recur at times $\{T_0,
	\dots, T_k\} = \ker(x-x(0)) \cup \{T\}$. Since time series from real-world
measurements contain noise, we consider the time series to be
$\varepsilon$-$\delta$-approximately recurring, as
in \cref{def:approx2}.
By measuring the Euclidean distance of each point in the time series from the
starting point $x(0)$, we construct a scalar-valued function $v_x: I \to [0,\infty )$, defined by $v_x(t) := \|x(t) - x(0)\|_2$ mapping each time point to its
distance from the starting point. We then apply the sublevel set filtration to
$v_x$ to construct the persistence diagram $D(v_x)$.
According to \cref{def:approx2}, we are interested in the local
minima of $v_x$ within $\varepsilon$ from the starting point $x(0)$ and
are separated by a local maximum at least $\delta$ higher than the
minima themselves. To identify these significant local minima in the
persistence diagram $D(v_x)$, we consider points $(b, d)$ in $D(v_x)$ where $b <
	\varepsilon$ and $d - b > \delta$.
By mapping from the points in the persistence diagram $D(v_x)$ corresponding to
these significant local minima back to the time domain, we then obtain the
times of recurrence $\{T_0, \dots, T_k\}$.

A visualization of the method is provided in the first column of \cref{fig:method_viz}.

\ifNotArXiv
	\paragraph{\textbf{Stability.}}
	It can be shown that
	$\|v_x - v_{x'}\|_\infty \leq 2\|x - x'\|_\infty$ for two time series $x, x'$. Furthermore, we
	know by~\cite{Cohen2005} that $d_B(D(v_x), D(v_{x'})) \leq \|v_x -
		v_{x'}\|_\infty$ and thus
	\begin{align}
		d_B(D(v_x), D(v_{x'})) \leq 2\|x - x'\|_\infty.
	\end{align}
	This result generally holds for any metric space $(\mathbb{R}^m, v)$.
\fi

\paragraph{\textbf{Runtime Complexity.}}
The construction of $v_x$ requires $O(n)$ operations,
while the sorting procedure for the sublevel set filtration requires $O(n \log n)$,
where $n$ is the number of samples of $x$.
This yields a total runtime complexity of $O(n \log n)$.
\paragraph{\textbf{Applicability and Limitations.}}
This method is effective for analyzing recurring, repetitive, and periodic time
series without self-intersections in state space, even in the presence of noise
and non-uniform sampling conditions. However, it is unsuitable for time series that fail
to return to a point near the starting point due to trends.

\subsection{Delay Embeddings for Repetitive Series (Method 2)}

\paragraph{\textbf{Method.}} When a time series $x$ self-intersects at times $t_1, t_2$ where
$x(t_1) = x(t_2)$, and particularly when $t_1 = 0$, our methods may falsely identify
$[t_1, t_2]$ as a repetition. This can be resolved either through delay embedding
(if trajectories around $t_1$ and $t_2$ differ over some neighborhood covered by
the embedding) or through differentiation~\cite{Bauer2024}, though the latter is
more sensitive to noise and dimensional scaling.
Therefore, we utilize the delay embedding series $U_x \colon I
	\to \nR^{md}, U_x(t) = (x(t), x(t+\Delta), \dots, x(t+(d-1)\Delta))$ of the time series $x$
with embedding dimension $d$ and time delay $\Delta$ to construct a scalar-valued
function $v_x(t) = \|U_x(t) - U_x(0)\|_p$. We then apply the sublevel set
filtration to $v_x$ to identify the times of recurrence as described in the
previous method.

\ifNotArXiv
	\paragraph{\textbf{Stability.}}
	Given the delay embeddings $U_x, U_{x'}$
	of two time series $x, x'$, it can be shown that
	\begin{equation}
		d_B(D(v_x), D(v_{x'})) \leq  \|v_x - v_{x'}\|_\infty \leq 2\sqrt[p]{d} \|x - x'\|_\infty
	\end{equation}
\fi

\paragraph{\textbf{Runtime Complexity.}}
The constructions of $U_x$ and then of $v_x$ respectively require $O(n)$ operations,
while the sorting procedure for the sublevel set filtration requires $O(n \log n)$,
where $n$ is the number of samples of $x$.
This yields a total runtime complexity of $O(n \log n)$.
\paragraph{\textbf{Applicability and Limitations.}}
This method is applicable to self-intersecting recurrent and repetitive time
series and is robust against noise and non-uniform sampling. However, it is not
suitable for time series not returning to a point near the start
due to trends.
Compared to the previous method, this approach is more sensitive to noise
due to the higher dimensionality of the delay embedding and requires
a suitable choice of embedding dimension $d$ and time delay $\Delta$.

\subsection{Recurrence Functions for Periodic Series (Method 3)}
\paragraph{\textbf{Method.}} While delay embeddings avoid issues with self-intersections
at the starting point, they require careful parameter tuning. A simpler alternative is
to consider the recurrence function $w_x: I \times I \to [0,\infty )$ defined as
$w_x(t_1, t_2) = \|x(t_1) - x(t_2)\|_2$. Computing its diagonal averages
$v_{x}(\Delta) = \text{avg}_t w_x(t, t+\Delta)$ and applying the sublevel set
filtration yields another method to identify recurrence times.

\ifNotArXiv
	\paragraph{\textbf{Stability.}}
	Given the recurrence functions $w_x, w_{x'}$ of two time series $x, x'$ and the surrogates $v_{x}, v_{x'}$, using the reverse triangular inequality it can be shown that
	$\|v_{x} - v_{x'}\|_\infty  \leq  \|w_x - w_{x'}\|_\infty \leq 2 \|x - x'\|_\infty $. Together with
	the stability result from~\cite{Cohen2005} the whole chain of inequalities gives
	\begin{align}
		d_B(D(v_{x}), D(v_{x'})) \leq 2\|x - x'\|_\infty.
	\end{align}
\fi

\paragraph{\textbf{Runtime Complexity.}}
The construction of $w_{x}$  requires $O(n^2)$ operations. If we use
$w_{x}(t_1, t_2) = \|x(t_1) - x(t_2)\|_2^2$ this can be sped up to $O(n\log n)$
by evaluating two cumulative sums ($O(n)$) and one autocorrelation ($O(n\log
	n)$), while the sorting procedure for the sublevel set filtration requires $O(n
	\log n)$, where $n$ is the number of samples of $x$. This yields a total
runtime complexity of $O(n \log n)$.

\paragraph{\textbf{Applicability and Limitations.}}
The averaging method has both strengths and limitations. While it is robust to noise and non-uniform sampling, it primarily detects $\Delta$-periodic behavior rather than general recurrence. It requires $x(t) \approx x(t + \Delta)$ to hold consistently for effective detection, and requires careful implementation for non-uniform sampling.

\section{Experiments}

\subsection{Dataset}

Since this is the first work on the estimation of cycle lengths in
multi-variate time series with periodic, repetitive, and recurring behavior,
we introduce a new dataset comprising a multi-variate time series displaying
these characteristics.
To this end, we utilize a partially simulated industrial environment
combining real industrial control system hardware with accurate simulations of the underlying physical processes within an injection molding machine.

The first part of the time series (I.1: Cycles 1-20 and I.2: Cycles 21-40)
exhibits periodic behavior, achieved by consistently repeating the same
production process without altering system parameters.

Following the initial dataset segments, Part II introduces variations in the
timing of production phases
resulting in an
(approximately) repetitive state space trajectory following the same path in
state space as the standard procedure but at different speeds.
Please refer to \cref{tab:dataset_config} for the specific configurations
of each section. Each variant simulates practical
deviations possible in standard production cycles.

Part III introduces higher friction in the mechanical movements during the
production sequence, resulting in deviating sensor measurements and a recurring
state space trajectory. This part is designed to enforce the divergence from
the state space trajectory of the standard production sequence. III.1.1
involves increased friction on the plastification axis; III.1.2 does the same
on the injection axis; and III.1.3 combines these frictions.

The subsections of
I.2, II.2, and III.2 replicate the conditions of
I.1, II.1, III.1 respectively
but add uniformly distributed measurement noise to assess the
robustness of our methods.

Each of these scenarios is designed to test the limits of our methods and to
provide a comprehensive benchmark for evaluating the effectiveness of future approaches in repetition time estimation in a realistic industrial setting.

\cref{tab:dataset_config} gives a comprehensive overview about the
different sections of the time series, their type of behavior as well
as the measures taken in simulation.

\begin{table}[ht]
	\centering
	\caption{Overview of dataset configurations}
	\label{tab:dataset_config}
	\begin{tabular}{lllp{6cm}}
		\toprule
		Cycles  & Section & Behavior   & Description                                                  \\
		\midrule
		1-20    & I.1     & Periodic   & Standard production sequence                                 \\
		21-40   & I.2     & Periodic   & I.1 with added measurement noise                             \\
		41-45   & II.1.1  & Repetitive & 1-second dead time at ejector                                \\
		46-50   & II.1.2  & Repetitive & Slower injection speed                                       \\
		51-55   & II.1.3  & Repetitive & Slower plastification process                                \\
		56-60   & II.1.4  & Repetitive & Delayed clamping sequence                                    \\
		61-65   & II.1.5  & Repetitive & Delay in the ejector's operation                             \\
		66-70   & II.2.1  & Repetitive & II.1.1 with measurement noise                                \\
		71-75   & II.2.2  & Repetitive & II.1.2 with measurement noise                                \\
		76-80   & II.2.3  & Repetitive & II.1.3 with measurement noise                                \\
		81-85   & II.2.4  & Repetitive & II.1.4 with measurement noise                                \\
		86-90   & II.2.5  & Repetitive & II.1.5 with measurement noise                                \\
		91-95   & III.1.1 & Recurring  & Increased friction on the plastification axis                \\
		96-100  & III.1.2 & Recurring  & Increased friction on the injection axis                     \\
		101-105 & III.1.3 & Recurring  & Increased friction on both plastification and injection axes \\
		106-110 & III.2.1 & Recurring  & III.1.1 with measurement noise                               \\
		111-115 & III.2.2 & Recurring  & III.1.2 with measurement noise                               \\
		116-120 & III.2.3 & Recurring  & III.1.3 with measurement noise                               \\
		\bottomrule
	\end{tabular}
\end{table}

\subsection{Evaluation procedure}
Consider a finite time series $x(t_1), \dots, x(t_n)$, observed at
$t_1 < \dots < t_n$ and spanning \(m\) dimensions featuring \enquote{cyclic
	behavior} as defined in \cref{sec:definitions}. This series comprises a
total of $k$ cycles, with the corresponding recurrence times denoted by $T_1
	\ldots, T_k$. To assess the precision of our
methods estimates $\hat{T}_1, \ldots,
	\hat{T}_k$, we introduce the following evaluation procedure, which can be
reused for future research involving the proposed dataset. We will adhere to
this procedure to independently test each of the three methods we
introduced.
To explore the varying strengths and weaknesses of each method based on the
data characteristics, we propose to evaluate each method independently on
sections I, II, and III of the data. This evaluation will be conducted
separately on segments of the time series without added measurement noise (I.1,
II.1.1-5, III.1.1-3) and on those with measurement noise.

Furthermore, we aim to assess the overall effectiveness of each method across
the entire dataset, encompassing all variations and noise conditions. This
allows to evaluate the robustness and generalization capabilities of each method
across different types of cyclic behaviors and noise levels.

Let \(\tau = [\tau_1,\ldots,\tau_k]\) and \(\hat{\tau} = [\hat{\tau}_1,\ldots,\hat{\tau}_k]\) be the true and estimated cycle lengths respectively. We evaluate using the mean absolute error (MAE) and mean absolute relative error (MARE) metrics, where the MARE accounts for varying cycle length magnitudes.

\begin{align*}
	\operatorname{MAE}(\tau, \hat{\tau})   = \frac{1}{k} \sum_{i=1}^k |\tau_i - \hat{\tau}_i| \hspace{1cm}
	\operatorname{MARE}(\tau, \hat{\tau})  = \frac{1}{k} \sum_{i=1}^k \left| \frac{\tau_i - \hat{\tau}_i}{\tau_i} \right|
\end{align*}

\subsection{Results}

\begin{figure}
	\begin{minipage}{\textwidth}
		\centering
		\ifNotArXiv \vspace{-0.75cm} \fi
		\includegraphics[width=0.95\textwidth]{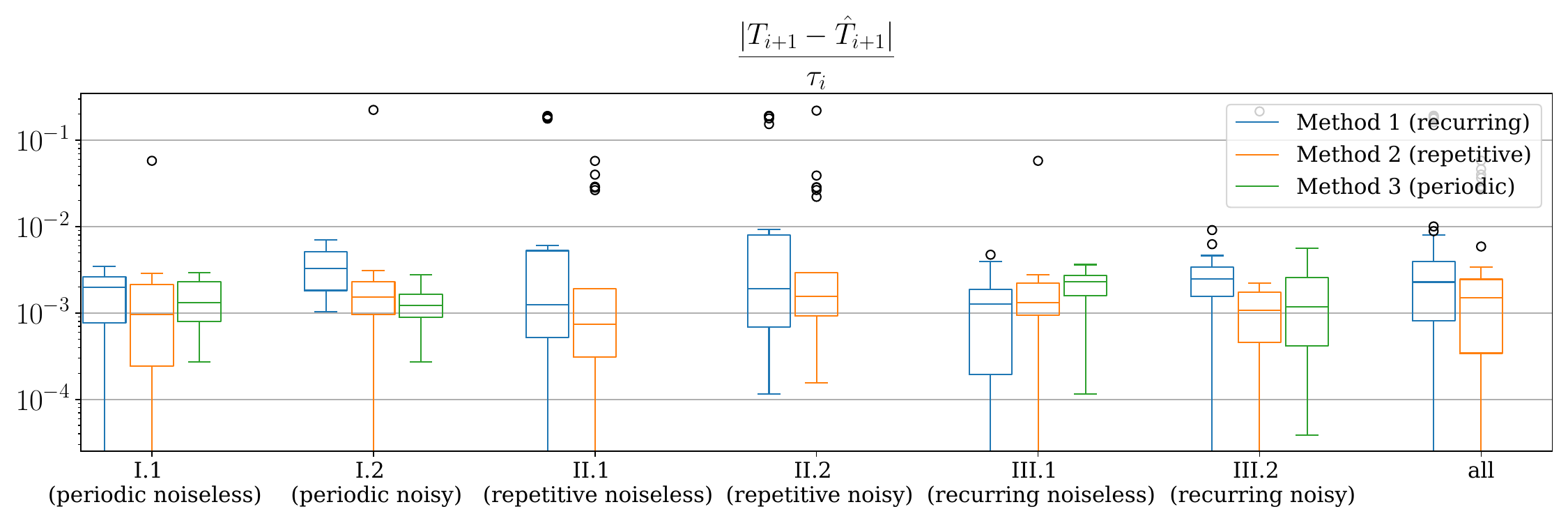}
		\caption{Distribtion of the absolute error of predicted recurrence times relative to the true cycle duration for each method and dataset section.}
		\label{fig:boxplot_T}
	\end{minipage}

	\begin{minipage}{\textwidth}
		\centering
		\vspace{0.25cm}
		\captionof{table}{Evaluation results with MAE measured in samples.}
		\label{tab:results}

		\begin{adjustbox}{max width=\textwidth}
			\begin{tabular}{@{}lrrrrrrrrrrrrrr@{}}
				\toprule
				\multirow{3}{*}{Method}                                                           &
				\multicolumn{4}{c}{I\,\scriptsize{(periodic)}}                                    &
				\multicolumn{4}{c}{II\,\scriptsize{(repetitive)}}                                 &
				\multicolumn{4}{c}{III\,\scriptsize{(recurring)}}                                 &
				\multicolumn{2}{c}{All sections}                                                                                                                                        \\
				\cmidrule(lr){2-5}  \cmidrule(lr){6-9}  \cmidrule(lr){10-13} \cmidrule(lr){14-15} &
				\multicolumn{2}{c}{I.1\,\scriptsize{(noiseless)}}                                 &
				\multicolumn{2}{c}{I.2\,\scriptsize{(noisy)}}                                     &
				\multicolumn{2}{c}{II.1\,\scriptsize{(noiseless)}}                                &
				\multicolumn{2}{c}{II.2\,\scriptsize{(noisy)}}                                    &
				\multicolumn{2}{c}{III.1\,\scriptsize{(noiseless)}}                               &
				\multicolumn{2}{c}{III.2\,\scriptsize{(noisy)}}                                   &
				\multicolumn{2}{c}{}                                                                                                                                                    \\
				                                                                                  & \multicolumn{1}{c}{~MAE~} & \multicolumn{1}{c}{~MARE~~}
				                                                                                  & \multicolumn{1}{c}{~MAE~} & \multicolumn{1}{c}{~MARE~~}
				                                                                                  & \multicolumn{1}{c}{~MAE~} & \multicolumn{1}{c}{~MARE~~}
				                                                                                  & \multicolumn{1}{c}{~MAE~} & \multicolumn{1}{c}{~MARE~~}
				                                                                                  & \multicolumn{1}{c}{~MAE~} & \multicolumn{1}{c}{~MARE~~}
				                                                                                  & \multicolumn{1}{c}{~MAE~} & \multicolumn{1}{c}{~MARE~~}
				                                                                                  & \multicolumn{1}{c}{~MAE~} & \multicolumn{1}{c}{~MARE}                               \\
				\cmidrule(r){1-1} \cmidrule(lr){2-3} \cmidrule(lr){4-5} \cmidrule(lr){6-7} \cmidrule(lr){8-9} \cmidrule(lr){10-11} \cmidrule(lr){12-13} \cmidrule(l){14-15}

				1\,\scriptsize{(recurring)}                                                       & \bf{23.35}                & \bf{0.001}~                 & 55.95       & 0.002~
				                                                                                  & 605.20                    & 0.020~                      & 632.72      & 0.021~
				                                                                                  & 52.07                     & 0.002~                      & 62.79       & 0.002~
				                                                                                  & 287.08                    & 0.010                                                   \\
				\scriptsize{$\epsilon=0.3, \delta=0.6$}                                                                                                                                 \\ \midrule

				2\,\scriptsize{(repetitive)}                                                      & 94.50                     & 0.004~                      & 326.85      & 0.013~
				                                                                                  & \bf{163.28}               & \bf{0.006}~                 & \bf{350.60} & \bf{0.013}~
				                                                                                  & 114.53                    & 0.004~                      & 427.64      & 0.017~
				                                                                                  & \bf{80.13}                & \bf{0.003}                                              \\
				\scriptsize{$\epsilon=0.4, \delta=0.5$}                                                                                                                                 \\
				\scriptsize{$d=4, \tau=500$}                                                                                                                                            \\ \midrule

				3\,\scriptsize{(periodic)}                                                        & \bf{23.35}                & \bf{0.001}~                 & \bf{32.4}   & \bf{0.001}~
				                                                                                  & -                         & -~                          & -           & -~
				                                                                                  & \bf{32.67}                & \bf{0.001}~                 & \bf{41.21}  & \bf{0.002}~
				                                                                                  & -                         & -                                                       \\
				\scriptsize{$\epsilon=0.7, \delta=0.1$}                                                                                                                                 \\ \bottomrule
			\end{tabular}
		\end{adjustbox}
		\vspace{-0.5cm}
	\end{minipage}
\end{figure}

\cref{tab:results} summarizes the MAE and MARE for each method and dataset
section (best results in bold), while \cref{fig:boxplot_T} shows the
distribution of the MAE of estimated times of recurrence. For cases where a
method failed to identify the correct number of cycles, results are omitted
(marked with dashes in the table and excluded from the boxplots).

The experimental results demonstrate the relative efficacy and effectiveness of each method
across different data characteristics.
Method 3 excels in detecting periodic behavior (Section I, cycles 1-40),
particularly under real-world conditions (cycles 21-40), due to its
noise-mitigating averaging mechanism. It also performs best with recurring
behavior showing stable cycle times (Section III, cycles 91-120). However, it
fails to detect repetitive behavior with varying cycle lengths (Section II,
cycles 41-90) due to the inherent assumption of constant cycle lengths.

Method 2 achieves the best overall performance, particularly excelling in
Section II through its effective delay embedding approach. However, it shows
higher noise sensitivity due to its increased dimensionality, aligning with our
theoretical analysis\ifArXiv\ in
	\cref{sec:stability_proofs}\fi.

Method 1's simpler approach outperforms Method 2 in Sections I and III, though
not matching Method 3's accuracy. It underperforms in Section II, where Method
2's specialized design proves more effective.

\section{Discussion and conclusion}

This paper introduced a novel and computationally efficient ($O(n \log n)$) framework for identifying repetitions in
multi-variate time series using persistent homology, along with three
specialized methods tailored to different types of cyclic behavior. The
experimental results on our benchmark dataset demonstrate both the
effectiveness and limitations of each method, providing valuable insights for
practical applications.

Method 3, which averages over constant delays, proves particularly
effective for periodic time series under both conditions with and without noise, which can be attributed to
the averaging mechanism filtering out zero-mean measurement
noise. However, its performance deteriorates significantly when handling
repetitive patterns with varying cycle lengths.
Method 2, utilizing delay embeddings, demonstrates superior performance across
all sections combined and particularly excels in handling
repetitive patterns. The method's ability to capture the temporal evolution
through delay embeddings enables it to effectively distinguish between
different segments of the time series. However, this comes at the cost of
increased sensitivity to noise, as evidenced by the larger increase in MARE
under noisy conditions.
Method 1, while being the most general and straightforward approach, shows
robust performance in both periodic and recurring sections but is outperformed
by Method 2 in the repetitive regime.

The stability guarantees of all three methods provide a theoretical
foundation for their reliability, while the experimental results validate their
practical applicability. The complementary strengths of the methods suggest
the choice of method to be guided by the expected characteristics of
the time series and the specific requirements of the application.

Our high quality benchmark dataset
\footnote{Dataset and code: \url{https://github.com/JRC-ISIA/paper-2025-idsc-topology-driven-identification-of-repetitions-in-multi-variate-time-series}}
from an industrial setting demonstrates multiple generalizations of cyclical behavior, providing a valuable resource for future research in this area.

Future work could explore designing specialized methods within this framework
for other types of cyclic behavior, such as quasi-periodic or chaotic time
series.

In conclusion, this work advances both the theoretical understanding of
topology-based cycle detection in multi-variate time series and provides
practical tools for analyzing real-world data, particularly in industrial
applications, where robust cycle detection is crucial for monitoring and control
tasks.

\ifNotBlindReview
	\section*{Acknowledgment}
	The financial support by the Christian Doppler Research Association, the
	Austrian Federal Ministry for Digital and Economic Affairs and the Federal
	State of Salzburg is gratefully acknowledged.
\fi

\printbibliography

@incollection{White2001,
  title = {Transforms, Wavelets},
  editor = {S. Braun},
  booktitle = {Encyclopedia of Vibration},
  publisher = {Elsevier},
  address = {Oxford},
  pages = {1419-1435},
  year = {2001},
  isbn = {978-0-12-227085-7},
  author = {P. White},
}

@inproceedings{Huber2020,
  title = {Persistent Homology in Data Science},
  author = {Stefan Huber},
  booktitle = {Proc.\ 3rd Int.\ Data Science Conference (iDSC'20)},
  series = {Data Science -- Analytics and Applications},
  year = 2020,
  month = may,
  address = {Dornbirn, Austria (virtual)},
  doi = {10.1007/978-3-658-32182-6_13},
}

@article{Bonis2024,
  title = {Topological phase estimation method for reparameterized periodic
           functions},
  author = {Bonis, Thomas and Chazal, Fr{\'e}d{\'e}ric and Michel, Bertrand and
            Reise, Wojciech},
  journal = {Advances in Computational Mathematics},
  volume = {50},
  number = {4},
  pages = {66},
  year = {2024},
  publisher = {Springer},
}

@inproceedings{Perea2016,
  title = {Persistent homology of toroidal sliding window embeddings},
  booktitle = {2016 {IEEE} {International} {Conference} on {Acoustics}, {Speech}
               and {Signal} {Processing} ({ICASSP})},
  publisher = {IEEE},
  author = {Perea, Jose A},
  year = {2016},
  pages = {6435--6439},
}

@article{Perea2015,
  title = {{SW1PerS}: {Sliding} windows and 1-persistence scoring; discovering
           periodicity in gene expression time series data},
  volume = {16},
  journal = {BMC bioinformatics},
  author = {Perea, Jose A and Deckard, Anastasia and Haase, Steve B and Harer,
            John},
  year = {2015},
  note = {Publisher: Springer},
  pages = {1--12},
}

@article{Perea2019,
  title = {Topological time series analysis},
  volume = {66},
  number = {5},
  journal = {Notices of the American Mathematical Society},
  author = {Perea, Jose A},
  year = {2019},
  note = {Publisher: American Mathematical Society, AMS},
  pages = {686--694},
}

@article{Perea2015a,
  title = {Sliding {Windows} and {Persistence}: {An} {Application} of {
           Topological} {Methods} to {Signal} {Analysis}},
  volume = {15},
  issn = {1615-3383},
  url = {https://doi.org/10.1007/s10208-014-9206-z},
  doi = {10.1007/s10208-014-9206-z},
  number = {3},
  journal = {Foundations of Computational Mathematics},
  author = {Perea, Jose A. and Harer, John},
  month = jun,
  year = {2015},
  pages = {799--838},
}

@article{Perea2018,
  title = {(Quasi) periodicity quantification in video data, using topology},
  author = {Tralie, Christopher J and Perea, Jose A},
  journal = {SIAM Journal on Imaging Sciences},
  volume = {11},
  number = {2},
  pages = {1049--1077},
  year = {2018},
  publisher = {SIAM},
}

@article{Ichinomiya2023,
  title = {Time series analysis using persistent homology of distance matrix},
  author = {Ichinomiya, Takashi},
  journal = {Nonlinear Theory and Its Applications, IEICE},
  volume = {14},
  number = {2},
  pages = {79--91},
  year = {2023},
  publisher = {The Institute of Electronics, Information and Communication
               Engineers},
}

@article{Yang2016,
  title = {Time-domain period detection in short-duration videos},
  volume = {10},
  url = {https://doi.org/10.1007/s11760-015-0797-x},
  doi = {10.1007/s11760-015-0797-x},
  number = {4},
  journal = {Signal, Image and Video Processing},
  author = {Yang, Jing and Zhang, Hong and Peng, Guohua},
  year = {2016},
  pages = {695--702},
}

@misc{Bauer2024,
  title = {Cycling {Signatures}: {Identifying} {Cycling} {Motions} in {Time} {
           Series} using {Algebraic} {Topology}},
  url = {https://arxiv.org/abs/2312.04734},
  author = {Bauer, Ulrich and Hien, David and Junge, Oliver and Mischaikow,
            Konstantin},
  year = {2024},
  note = {\_eprint: 2312.04734},
}

@inproceedings{Cohen2005,
  title = {Stability of persistence diagrams},
  author = {Cohen-Steiner, David and Edelsbrunner, Herbert and Harer, John},
  booktitle = {Proceedings of the twenty-first annual symposium on Computational
               geometry},
  pages = {263--271},
  year = {2005},
}

@misc{Mueller2025,
  title = {Open Challenges in Time Series Anomaly Detection: An Industry
           Perspective},
  author = {Andreas Mueller},
  year = {2025},
  eprint = {2502.05392},
  archivePrefix = {arXiv},
  primaryClass = {cs.LG},
  url = {https://arxiv.org/abs/2502.05392},
}

@book{Edelsbrunner2010,
  title = {Computational topology: an introduction},
  author = {Edelsbrunner, Herbert and Harer, John},
  year = {2010},
  publisher = {American Mathematical Soc.},
}

\ifArXiv
	\newpage
	\appendix

	\section{Stability Proofs}\label{sec:stability_proofs}
	In this section, we provide the stability proofs for the three methods introduced in \cref{sec:methods}.

	\subsection{Method 1}

	Let $x$ denote a time series $I \to \nR^m$, let $v_x \colon I \to \nR$ denote
	the distance function $v_x(t) = \|x(t) - x(0)\|_2$, and let $D(v_x)$ denote the
	persistence diagram of the sublevel-set filtration of $v_x$. The claim is that
	close time series $x, x'$ in terms of the supremum norm lead to close
	persistence diagrams $D(v_x), D(v_{x'})$ in terms of the bottleneck distance
	$d_B$, i.e., the composed map $x \mapsto v_x \mapsto D(v_x)$ is Lipschitz.
	(Note that $\nR^m$ is endowed with the 2-norm here, but of course other
	$p$-norms can be used, too. In particular, $\|x\|_\infty = \sup_{t \in I}
		\|x(t)\|_2$.)

	Let $x, x'$ be two time series $I \to \nR^m$ then for each $t \in I$ we have
	\begin{align*}
		|v_x(t) - v_{x'}(t)| & = \|x(t) - x(0) - x'(t) + x'(0)\|_2          \\
		                     & \leq \|x(t) - x'(t)\|_2 + \|x(0) - x'(0)\|_2 \\
		                     & \leq 2\|x - x'\|_\infty
	\end{align*}
	and hence $\|v_x - v_{x'}\|_\infty \leq 2\|x - x'\|_\infty$. Furthermore, we
	know by~\cite{Cohen2005} that $d_B(D(v_x), D(v_{x'})) \leq \|v_x -
		v_{x'}\|_\infty$ and thus
	\begin{align}
		d_B(D(v_x), D(v_{x'})) \leq 2\|x - x'\|_\infty.
	\end{align}

	This result generally holds for any metric space $(\mathbb{R}^m, v)$ as long as
	the distance function $v$ agrees with the reverse triangle inequality, i.e.,
	$|v(x) - v(y)| \leq |v(x - y)|$ for all $x, y \in \mathbb{R}^m$.

	\subsection{Method 2}

	Method 2 generalized method 1 by adding a time delay embedding $U_x \colon I
		\to \nR^{md}, U_x(t) = (x(t), x(t+\tau), \dots, x(t+(d-1)\tau))$ to our
	consideration, i.e., we consider the composed map $x \mapsto U_x \mapsto v_x
		\mapsto D(v_x)$. Note that $U_x(t)$ can itself be considered a time series in
	$I \to \nR^{md}$. In particular, when $d=1$ then method 2 equals method 1
	again.
	As before, we claim that  close time series $x, x'$ in terms of the supremum
	norm lead to close persistence diagrams $D(v_x), D(v_{x'})$ in terms of the
	bottleneck distance $d_B$, i.e., the composed map $x \mapsto \cdots \mapsto
		D(v_x)$ is Lipschitz again.

	Let $x, x'$ be two time series $I \to \nR^m$ then for each $t \in I$ we have
	\begin{align*}
		|v_x(t) - v_{x'}(t)| & = \|U_x(t) - U_x(0) - U_{x'}(t) + U_{x'}(0)\|_2          \\
		                     & \leq \|U_x(t) - U_{x'}(t)\|_2 + \|U_x(0) - U_{x'}(0)\|_2 \\
		                     & \leq 2\|U_x - U_{x'}\|_\infty.
	\end{align*}
	Furthermore note that
	\begin{align*}
		\|U_x(t)\|^2 = \sum_{i=1}^d \|x(t + (i-1)\tau)\|_2^2 \leq d \|x\|_\infty^2
	\end{align*}
	and hence $\|U_x - U_{x'}\|_\infty \leq \sqrt{d} \|x - x'\|_\infty$, leading to
	\begin{align}
		d_B(D(v_x), D(v_{x'})) \leq 2\sqrt{d} \|x - x'\|_\infty.
	\end{align}

	\subsection{Method 3}
	In Method 3, we take a time series $x \colon I \mapsto \nR^m$, construct the
	function $w_x \colon I \times I \to \mathbb{R}$ from $x$ as $w_x(t_1,t_2) =
		\|x(t_1) - x(t_2)\|_2$ and then average over diagonals of $I^2$, i.e., we
	define $v_{x}(\Delta) = \avg_t w_x(t,t+\Delta)$. We then apply the sublevel set
	filtration to $v_{x}$, i.e., we consider the composed map $x \mapsto w_x \mapsto
		v_{x} \mapsto D(v_{x})$.

	Similar to Methods 1 and 2, we claim that close time series $x, x'$ in terms of
	the supremum norm lead to close persistence diagrams $D(v_{x}), D(v_{x'})$ in
	terms of the bottleneck distance $d_B$, i.e., the composed map $x \mapsto
		\cdots \mapsto D(v_{x})$ is Lipschitz.

	Let $x, x'$ be two time series $I \to \nR^m$ then we have for all $t_1, t_2 \in
		I$
	\begin{align*}
		|w_x(t_1, t_2) - w_{x'}(t_1, t_2)| & = \left| \|x(t_1) - x(t_2)\|_2 - \|x'(t_1) - x'(t_2)\|_2 \right| \\
		                                   & \leq \|x(t_1) - x(t_2) - x'(t_1) + x'(t_2)\|_2                   \\
		                                   & \leq \|x(t_1) - x'(t_1)\|_2 + \|x(t_2) - x'(t_2)\|_2             \\
		                                   & \leq 2 \|x - x'\|_\infty
	\end{align*}
	using the reverse triangular inequality, and hence $\|w_x - w_{x'}\|_\infty
		\leq 2\|x - x'\|_\infty$. Next observe that for all $\Delta$ we have
	\begin{align*}
		|v_{x}(\Delta) - v_{x'}(\Delta)| & = |\avg_t w_x(t, t+\Delta) - \avg_{t'} w_{x'}(t', t'+\Delta)|    \\
		                                 & = |\avg_t \avg_{t'} w_x(t, t+\Delta) - w_{x'}(t', t'+\Delta)|    \\
		                                 & \leq \avg_t \avg_{t'} |w_x(t, t+\Delta) - w_{x'}(t', t'+\Delta)| \\
		                                 & \leq \|w_x - w_{x'}\|_\infty,
	\end{align*}
	and hence $\|v_{x} - v_{x'}\|_\infty \leq \|w_x - w_{x'}\|_\infty$. Together with
	the stability result from~\cite{Cohen2005}, i.e., $d_B(D(v_{x}), D(v_{x'})) \le
		\|v_{x} - v_{x'}\|_\infty$, the whole chain of inequalities gives
	\begin{align}
		d_B(D(v_{x}), D(v_{x'})) \leq 2\|x - x'\|_\infty.
	\end{align}

\fi

\end{document}